\documentclass{iau} 
\usepackage{graphicx}
\usepackage{natbib}

\pubyear{2022}
\volume{373}
\jname{Resolving the Rise and Fall of Star Formation in Galaxies}
\editors{Kim W.-T. \& Wong T., eds.}

\title[Evolution of And\,IX] 
{The formation and evolution of  Andromeda IX}

\author[Hedieh Abdollahi et al]   
{Hedieh Abdollahi$^1$, Atefeh Javadi$^2$, Mohammad Taghi Mirtorabi$^{1}$, Elham Saremi$^{2,3,4}$, Habib Khosroshahi$^2$, Jacco Th. van Loon$^5$, Iain McDonald$^{6,7}$, Elahe Khalouei$^2$, Sima T. Aghdam$^2$, \and Maryam Saberi$^8$}

\affiliation{Physics Department, Faculty of Physics and Chemistry, Alzahra University, Vanak, 1993891176, Tehran, Iran\\ email: {\tt Abdollahi.hedieh2013@gmail.com} \\[\affilskip]

	$^2$School of Astronomy, Institute for Research in Fundamental Sciences (IPM), Tehran, 19568-36613, Iran \\
	$^3$Instituto de Astrof{\`i}sica de Canarias, C/ V{\`i}a L{\`a}ctea s/n, 38205 La Laguna, Tenerife, Spain \\
	$^4$Departamento de Astrof{\`i}sica, Universidad de La Laguna, 38205 La Laguna, Tenerife, Spain\\
	$^5$Lennard-Jones Laboratories, Keele University, ST5 5BG, UK \\	$^6$Jodrell Bank Centre for Astrophysics, Alan Turing Building, University of Manchester, M13 9PL, UK\\
    $^7$Department of Physical Sciences, The Open University, Walton Hall, Milton Keynes, MK7 6AA, UK\\
	$^8$Rosseland Centre for Solar Physics, University of Oslo, P.O. Box 1029, Blindern, NO-0315, Oslo, Norway}

\begin{document}

\maketitle

\begin{abstract}
Local Group (LG), the nearest and most complete galactic environment, provides valuable information on the formation and evolution of the Universe. Studying galaxies of different sizes, morphologies, and ages can provide this information. For this purpose, we chose the And\,IX dSph galaxy, which is one of the observational targets of the Isaac Newton Telescope (INT) survey. 
A total of 50 long-period variables (LPVs) were found in And\,IX in two filters, Sloan $i'$ and Harris $V$ at a half-light radius of 2.5 arcmin. The And\,IX's star formation history (SFH) was constructed with a maximum star formation rate (SFR) of about $0.00082\pm0.00031$ M$\textsubscript{\(\odot\)}$ yr$^{-1}$, using LPVs as a tracer. The total mass return rate of LPVs was calculated based on the spectral energy distribution (SED) of about $2.4\times10^{-4}$ M$\textsubscript{\(\odot\)}$ yr$^{-1}$. The distance modulus of $24.56_{-0.15}^{+0.05}$ mag was estimated based on the tip of the red giant branch (TRGB).

\end{abstract}

\firstsection 
              
\section{Introduction} \label{sec:sec1}

Answering questions about the formation and evolution of the universe requires studying dwarf galaxies. While dwarf galaxies are the simplest system known, they are the most diverse in terms of star formation and chemical enrichment (\citealp{1998ARA&A..36..435M}).

A comprehensive survey (up to now) in the optical bands has been launched to explore the evolution of dwarf galaxies using the Isaac Newton Telescope (INT) (\citealp{Saremi2017, 2020ApJ...894..135S}). Probing long-period variable (LPV) stars, deriving the star formation history (SFH), and estimating the mass-loss of stars in galaxies are among the reasons for initiating the INT project. As a result of this survey, we can compare the SFHs of different galaxy types and study the evolution and quenching of dwarf galaxies (\citealt{golshan2017, Saremi2019, navabi2021}). The method to reconstruct SFH was first applied by \cite{javadi2011b} to rebuild the SFH of M\,33.

The asymptotic giant branch (AGB) stars were selected because they are at their most luminous stage ($\sim$ $10^4$ L\textsubscript{\(\odot\)} (\citealp{2005A&A...438..273V, Yuan2018})) and are easier to detect. AGBs are evolved stars and their luminosity is related to their birth mass (\citealp{Rezaei2014, hashemi2019}). Moreover, their high mass-loss ($10^{-7}$ $<$ $\dot{M}$ $\leq$ $10^{-3}$ M$\textsubscript{\(\odot\)}$ yr$^{-1}$ (\citealp{1999A&A...351..559V, javadi2013, javadi2019, Boyer2017})) contributes to enhancing the ISM.

The spheroidal dwarf galaxy And\,IX, located $\sim$ $39_{-2}^{+5}$ kpc (\citealp{2019MNRAS.489..763W}) along the major axis of M\,31 galaxy, was selected as our candidate for study.

\section{Photometry and detection of LPVs} \label{sec:sec2}
\subsection{Observations and photometry}
Nine observations were made from $2015-2017$ using the 2.5-m wide field camera (WFC) at INT in the Sloan $i'$, Harris $V$, and RGO $I$ filters. Data reduction is done using {\sc Theli} (Transforming Heavenly Light into Image), an image processing pipeline adapted to multi-chip cameras. The {\sc daophot} package performs the photometry procedures, and {\sc addstar} task can evaluate the completeness of the photometry (\citealp{1987PASP...99..191S}). Fully described details of the photometric procedure can be found in \cite{2020ApJ...894..135S}.

\subsection{Detection of LPVs as SFH tracers}
We detected LPVs using the method introduced by \cite{1996PASP..108..851S}. In this method, each star is assigned a variability index, which is a function of its magnitude and magnitude error. In each magnitude interval, a threshold for variability index is determined to separate variable stars from non-variables. More than $90\%$ of the stellar population in this magnitude range should have variability indexes higher than the threshold.

\begin{figure}[h]
\begin{center}
	\includegraphics[width=0.8\textwidth]{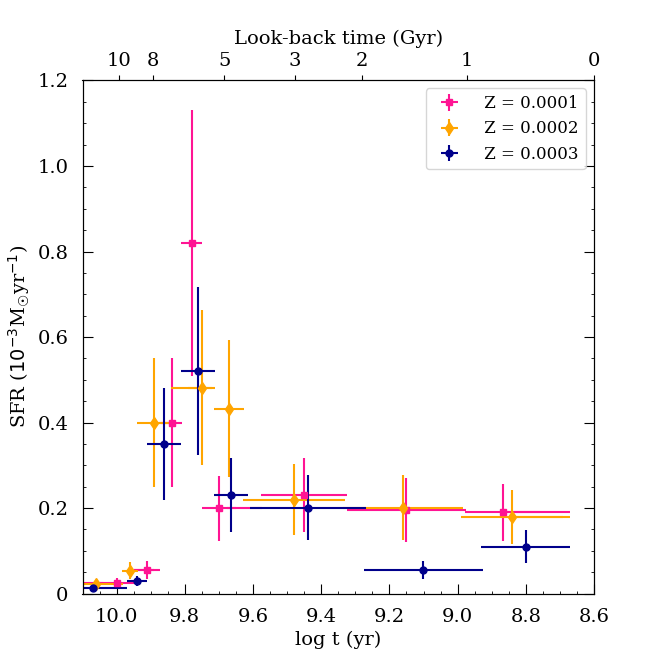}
	\caption{SFHs of And\,IX for metallicities of $Z = 0.0001$ (pink), $Z = 0.0002$ (orange), and $Z = 0.0003$ (blue) within two half-light radii ($\sim$ 0.022 deg$^2$).}
	\label{fig:fig1}
\end{center}
\end{figure}

To construct the SFH of And\,IX based on the LPVs, Milky Way foreground contamination should be excluded from candidates with an acceptable threshold. Omitting contamination was done by cross-correlating the INT catalog with the \emph{Gaia} DR$3$ (\citealp{Gaia2021}) and {\sc trilegal} simulation of Milky Way populations (\citealp{2005A&A...436..895G}).

We estimated the AGBs' birth mass through the birth mass-luminosity relation using PADOVA evolutionary tracks (\citealp{marigo2017}). AGBs are at the end of their evolutionary path, and their luminosity is a function of their birth mass (\citealp{javadi2017}). Age and pulsation duration are also calculated by mass-age and mass-pulsation relations (\citealp{javadi2011b, saremi2021}). The star formation rate (SFR) based on the mass, age, and pulsation duration over time represents the SFH. Fig.\ 1 presents SFH in two half-light radii (5 arcmin) of And\,IX in three metallicities. The highest peak of SFR reached at a level of $0.00082\pm0.00031$ M$\textsubscript{\(\odot\)}$ yr$^{-1}$ in 6 Gyr ago at Z = 0.0001. The metallicity of galaxies varies during their evolution, so in this study we consider metallicity $Z = 0.0002$ as well as a more metal-rich estimate $Z = 0.0003$
in addition to the main metallicity $Z = 0.0001$ (\citealp{2010MNRAS.407.2411C, 2012AJ....144....4M, 2013ApJ...779..102K, 2020ApJ...895...78W}).

\section{Probing of dust in And\,IX} \label{sec:sec3}
We modeled the spectral energy distribution (SED) by {\sc dusty} code to estimate the mass-loss ratio of LPVs (\citealp{1997}). {\sc dusty} assumes radiatively driven wind to solve hydrodynamic equations for AGB stars. Output parameters should be scaled for And\,IX as DUSTY solved equation by default parameters (gas-to-dust mass ratio $\psi\textsubscript{\(\odot\)}$ = 200, L = $10^4$ L$\textsubscript{\(\odot\)}$, and $\rho_{dust} = 3$ g $cm^{-3}$).

\section{Results and conclusions}
According to the different age gradients of the population in the inner and outer parts of the galaxy, the outside-in star formation scenario could be a galaxy evolution scenario.

The total mass return rate of our detected LPVs at Z = 0.0003, is estimated to be $\sim$ $10^{-4}$ M$\textsubscript{\(\odot\)}$ yr$^{-1}$.
By decreasing metallicity to Z = 0.0002, the total mass return increased by $50\%$. The total mass return rate is about $2.4\times10^{-4}$ M$\textsubscript{\(\odot\)}$ yr$^{-1}$ in And\,IX at Z = 0.0001. The average mass-loss rate by C-rich stars is more than $80\%$ of the total mass return rate in three metallicities.

Following are papers that attempt to estimate the total mass return rate of the variables and the SFH observed by the INT survey (\citealp{Saremi2017, 2020ApJ...894..135S}) for other dwarf galaxies.
\def\apj{{ApJ}}    
\def\nat{{Nature}}    
\def\jgr{{JGR}}    
\def\apjl{{ApJ Letters}}    
\def\aap{{A\&A}}   
\def\mnras{{MNRAS}}
\def\aj{{AJ}}
\let\mnrasl=\mnras

\end{document}